\documentstyle[12pt]{article}

\begin{document}
\begin{titlepage}

\title{The trace left by signature-change-induced compactification
         \thanks{
         Work supported by the Austrian Academy of Sciences
         in the framework of the ''Austrian Programme for
         Advanced Research and Technology''.}}

\author{Franz Embacher\\
        Institut f\"ur Theoretische Physik\\
        Universit\"at Wien\\
        Boltzmanngasse 5\\
        A-1090 Wien\\
        \\
        E-mail: fe@pap.univie.ac.at\\
        \\
        UWThPh-1994-55\\
        gr-qc/9411028}
\date{}

\maketitle

\begin{abstract}
Recently, it has been shown that an infinite succession of classical
signature changes (''signature oscillations'') can compactify and
stabilize internal dimensions, and simultaneously leads, after
a coarse graining type of average procedure, to an effective
(''physical'') space-time geometry displaying the usual Lorentzian
metric signature.
Here, we consider a minimally coupled scalar field on such an
oscillating background and study its effective dynamics. It turns out
that the resulting field equation in four dimensions contains a
coupling to some non-metric structure, the imprint of the
''microscopic'' signature oscillations on the effective properties of
matter. In a multidimensional FRW model, this structure is identical
to a massive scalar field evolving in its homogeneous mode.
\end{abstract}

\end{titlepage}

\section{Introduction}

There are various schemes according to which some of the space-time
dimensions compactify to unobservably small scales, while the
remaining ones evolve towards cosmologically reasonable sizes.
There are (although less) schemes in which the compactification of
internal dimensions occurs in a stable way, i.e. robust against small
perturbations (see e.g.
Refs. \cite{CremmerJulia}--\cite{Duff}).
Recently, one additional possibility to achieve this
has been added to the list \cite{FE2}. It relies on the
speculative possibility that the metric may change its
signature from Euclidean to Lorentzian type (and {\it vice versa})
in a {\it classical} process
\cite{EllisSumeruketal}--\cite{DrayHellaby}.
Each such process happens on a hypersurface whose intrinsic
geometries and extrinsic curvatures
(the latter describing the embedding into the full manifold)
coincide when computed from either side. Moreover, the
hypersurface is assumed to be spacelike with respect to the
Lorentzian side.
The classical signature change models should not be confused
with the mixed signature geometries appearing in
quantum cosmology \cite{GibbonsHartle}--\cite{Hayward2}, in which
case the corresponding hypersurfaces are required to have vanishing
extrinsic curvature.
\medskip

In Ref. \cite{FE2} (and in parts in Ref. \cite{FE1}),
this idea has been applied to a
Friedmann-Robertson-Walker (FRW) model with spatial sections
${\bf S}^3 \times {\bf S}^6$
and scale factors
$a_1$ and $a_2$, matter just being represented by a cosmological
constant $\Lambda$.
It turned out that there are solutions
$(a_1(\tau),a_2(\tau))$ such that the metric signature (i.e.
the signature of $g_{00}(\tau))$ oscillates rapidly. At large
values of the ''time'' coordinate $\tau$, the observable scale factor
$a_1$ grows linearly in $\tau$, while $a_2$ approaches the scale set
by $\Lambda^{-1/2}$ as its ''compactification radius''.
Postulating the effective (physically observable) metric as the
one obtained by an average over $g_{00}$ involving many oscillation
periods, the resulting space-time was found to inflate
exponentially in $a_1$.
\medskip

Considering the viability of such a scenario to describe the
actual universe, a lot of questions and problems arise. Some
of these have been posed (not answered) in Ref. \cite{FE2}.
Here, we would like to attack one such question, namely:
how does non-gravitational matter ''feel'' the underlying
signature oscillations? We will do a step towards
an answer by choosing a minimally coupled scalar field $\phi$ with
a self-interaction potential $V(\phi)$ as matter.
This provides a first orientation,
and we will leave the inclusion of higher spin fields
(in particular the interesting case of fermions) to future work.
Furthermore, we will consider the scalar field dynamics on
a background metric of the type described above. This is
easier than including the full back-reaction at the
fundamental level, and will even help us exhibiting the
structure that emerges.
\medskip

In Section 2, we will briefly review the scenario that has
been worked out in Ref. \cite{FE2}.
Thereafter, in Section 3, we define what the dynamics
of the scalar field at the fundamental level shall be,
and perform a coarse graining type of average in order
to obtain the effective field equation. The structure
emerging is a standard scalar field equation, supplemented
by a term $\tau\partial_\tau\phi$. In Section 4, this correction
is expressed in terms of a scalar $f$ which traces the
hypersurfaces of signature change in the effective geometry.
The result is a particular coupling of $\phi$ to the background
scalar $f$. Formaly, $f$ turns out to satisfy the
massive scalar field equation with
$m_f^2 = 2\Lambda/11$, although it is not clear to what extent
it develops features of a standard scalar field. Some of the structures
appearing here are likely to carry over to less symmetric situations,
as well as to models containing a richer matter sector.
Finally, in Section 5, we try to point out reasonable directions
for future research.

\section{Review of signature-change-induced compactification}
\setcounter{equation}{0}

Let us collect some results from Ref. \cite{FE2} that are
essential for the purposes pursued here. As already mentioned
in the introduction, we consider a background metric of the
type
\begin{equation}
ds^2_{\rm true} = - s(\tau) d\tau^2 + a_1(\tau)^2 d\sigma_3^2 +
               a_2(\tau)^2 d\sigma_6^2,
\label{2.1}
\end{equation}
where $d\sigma_n^2$ is the metric on the round unit n-sphere
${\bf S}^n$ (one may, for large $\tau$, ignore the curvature
of ${\bf S}^3$ and thus approximate $d\sigma_3^2$ by a
flat metric). As $\tau \rightarrow \infty$, the dominant
behaviour is given by $a_1 \rightarrow C \tau$ and
$a_2 \rightarrow (15/\Lambda)^{1/2}$, the next order displaying
damped oscillations such that
$\delta a_1/a_1 \sim \delta a_2/a_2 \sim \Lambda^2 \tau^{-2}$
during each period. These oscillations are induced by an
infinite succession of signature changes such that
$g_{00}^{\rm true} \equiv - s = \pm 1$. Hence we have $s=1$ for the
Lorentzian and $s=-1$ for the Euclidean periods. In the former
case, $\tau$ is given by the cosmological proper time, in the
latter case by its Euclidean analogue. The coordinate values
at which a change of signature occurs are denoted by $\tau_j$,
the corresponding  interval sizes being
$\Delta\tau_j = \tau_j - \tau_{j-1}$ (j=1,2,3...).
\medskip

For large $j$ (and hence large $\tau_j$) the
mixed-signature Einstein field equations (with cosmological
constant, but otherwise vaccum) imply to the lowest relevant order
\begin{equation}
\Delta\tau_j = \sqrt{\frac{3}{2\Lambda}}\,
               \bigg(\frac{1}{2j}+(-)^j \frac{11}{12\, j^{3/2}}
               \bigg),
\label{2.2}
\end{equation}
where Lorentzian periods correspond to even $j$, and
Euclidean ones to odd $j$. There is a slight predominance of the
Lorentzian over the Euclidean periods. Applying an appropriate
average (coarse graining) to the metric (\ref{2.1}) we found
that a small part of $g_{00}$ survives and gives rise to
the effective value
\begin{equation}
g_{00}^{\rm eff}(\tau) = - <s(\tau)>\, = - \,\frac{11}{\Lambda \tau^2}.
\label{2.3}
\end{equation}
The physical ten-metric resulting from this procedure is thus,
for large $\tau$,
\begin{equation}
ds^2_{\rm eff} = - \,\frac{11}{\Lambda} \frac{d\tau^2}{\tau^2}
                 + C^2 \tau^2 d\sigma_3^2
                 + \frac{15}{\Lambda} d\sigma_6^2.
\label{2.4}
\end{equation}
After a transformation of the time coordinate
\begin{equation}
\sqrt{\frac{11}{\Lambda}} \ln\bigg(\frac{\tau}{\tau_0}\bigg) =
     \eta,
\label{2.5}
\end{equation}
it takes the form
\begin{equation}
ds^2_{\rm eff} = - d\eta^2 +
                  \frac{K}{\Lambda} \exp\bigg(
                    2\sqrt{\frac{\Lambda}{11}}\,\eta\bigg)
                   d\sigma_3^2 +
                   \frac{15}{\Lambda} d\sigma_6^2,
\label{2.6}
\end{equation}
where $K$ can be given any value by a suitable choice of $\tau_0$.
The physical four-metric is obtained by omitting the
$d\sigma_6^2$-contribution in (\ref{2.4}) or (\ref{2.6}).
It nicely displays inflationary expansion, although it must be
said that a way out of this behaviour is not provided by the
simple model we are considering. This and other problems have
been discussed in Ref. \cite{FE2}. The general scenario of
compactification by signature change is likely to carry over
to a large class of topologies, the major condition being that
the internal space has non-zero Ricci curvature (and thus must
be at least two-dimensional).
\medskip

Let us at the end of this Section introduce the
abbreviation
\begin{equation}
f(\tau) \equiv \,<s(\tau)>\, = \frac{11}{\Lambda \tau^2} =
               \frac{11}{\Lambda\tau_0^2}
               \exp\bigg(-2\sqrt{\frac{\Lambda}{11}}\eta\bigg)
\label{2.7}
\end{equation}
that will be used in what follows and note that, whenever we
write $g_{00}^{\rm eff}$ or $g^{00}_{\rm eff}$, we refer to the metric
in the form (\ref{2.4}).

\section{Scalar field effective dynamics}
\setcounter{equation}{0}

Having a scheme at hand like the one described above,
a natural question is whether one may recover physical
laws that are compatible with our experience of
space, time and matter interactions. The general advantages
of the existence of compactified dimensions (a gauge
theory of elementary particles arising at least partially
from internal symmetries: see Ref. \cite{Duff} for a
recent review) are accompagnied here by the rather unusual
idea that within each second of physical time (as
measured by a clock) some underlying metric (namely
(\ref{2.1})) undergoes a great number of oscillations between
Lorentzian and Euclidean type, and only a small
predominance of the former is responsible for the
existence of a time evolution in the every-day
sense. Even if the oscillations may not be observable (due to
their extremely small time scales, and probably due to
quantum effects as well \cite{FE2}), one may ask how they
affect the properties of matter, and whether they produce
physical effects at the ''effective'' level. One such
effect is the inflationary behaviour of (\ref{2.6}) -- at
least in our simple model without any matter except a
cosmological constant.
\medskip

In order to proceed exploring
the consequences of the scenario we described, let us
study the behaviour of a minimally coupled real scalar field $\phi$.
By this we mean that the metric (\ref{2.1}) serves as a fixed
background structure at the ''true'' level. The aim is to
exhibit the dynamics of $\phi$ at the ''effective'' level
(which is characterized by the metric (\ref{2.4}), possibly without
the $d\sigma_6^2$-term). The effective dynamics of $\phi$
would thus be accessible to physical predictions and measurements.
Note that we neglect the back-reaction of the scalar field
onto the metric. This approximation is good enough to represent a
viable framework for our goals. The inclusion of $\phi$
into the full dynamics as well as the generalization to
various types of interacting matter fields are not expected to
provide drastic changes at the fundamental level at which the
question is posed: what is the trace left by the signature oscillations
in the observed world?
\medskip

In order to be specific, let $(x,y)$ denote the coordinates
on $({\bf S}^3,{\bf S}^6)$, $x^0 \equiv \tau$,
and let $\phi \equiv \phi(\tau,x,y)$.
The effect of the existence of internal dimensions on
observations in physical $(\tau,x)$-space is well known:
Expanding $\phi$ in terms of appropriate modes on ${\bf S}^6$,
one obtains an infinite ''tower'' of scalar fields
corresponding to the structure of possible internal
excitations (see e.g. Ref. \cite{BailinLove}). This is logically
independent of what we are aiming at, and may be performed at any
stage of our considerations (preferably at the end).
Hence, we will leave these issues aside, and simply allow
$\phi$ to depend on all coordinates, as already indicated
above. If one likes to study only the modes homogeneous
on ${\bf S}^6$, one sets $\phi \equiv \phi(\tau,x)$.
In even more drastic simplifications, one may let
$\phi \equiv \phi(\tau)$, as is usually done in minisuperspace
approaches to cosmology. The analysis given below will be
uneffected by such choices.
\medskip

The field equation at the ''true'' level shall be as simple
and natural as possible. Hence, we assume the fundamental
scalar field equation to be
\begin{equation}
\Box_{\rm true} \phi \equiv
g^{-1/2}\partial_\mu \,g^{1/2} g_{\rm true}^{\mu\nu}\partial_\nu\phi
= V'(\phi),
\label{3.1}
\end{equation}
where $g \equiv |\det(g_{\mu\nu}^{\rm true})|$ and $V$ is a
self-interaction potential. In the case of a free massive
field, one would specify $V(\phi) = m^2\phi^2/2$. This equation is
meant to apply for the Lorentzian and the Euclidean domains
separately. In other words, in (\ref{3.1})
$g_{\rm true}^{00} = -s(\tau)$ is considered as a constant
$\pm 1$. Denoting the spatial part of the metric by
$g_{ij}^{\rm true}$, and pulling the constant $s$ out of the
$\tau$-derivatives, we find
\begin{equation}
g^{-1/2}\partial_\tau \,g^{1/2}\partial_\tau \phi =
s \Big( g^{-1/2}\partial_i \,g^{1/2} g_{\rm true}^{ij}\partial_j\phi
- V'(\phi) \Big).
\label{3.2}
\end{equation}
This should be viewed as a second order time evolution equation.
The coefficients -- encoded in the $g_{\mu\nu}^{true}$ --
perform continuous damped oscillations, except for $s(\tau)$
which is now allowed to undergo an infinite succession of jumps
between $1$ and $-1$. The condition for the behaviour of
the scalar field at the matching hypersurfaces $\tau = \tau_j$
shall be contituity of $\phi$ and existence of
$\partial_\tau \phi$. As a consequence, $\partial_{\tau\tau} \phi$
becomes discontinuous, which is analogous to the evolution
of the scale factors \cite{FE2}. Note that, had we
retained $g_{\rm true}^{00}$ {\it inside} the
derivatives in (\ref{3.1}), an additional term containing
$\delta(\tau-\tau_j)$ would have emerged and prevented the
well-posed matching conditions associated with (\ref{3.2}).
This is in accordance with
Refs. \cite{EllisSumeruketal}--\cite{Ellis},
where a purely time-dependent
scalar is considered, and with
Refs. \cite{DrayManogueTucker1}--\cite{DrayManogueTucker2},
where the scalar field matching conditions are discussed in a
two-dimensional model. In Ref. \cite{DrayManogueTucker1},
by the way, it is shown that the jump conditions
imply particle production when $\phi$ is a quantum field.
(This is an aspect that seems worth being pursued further in the
present context as well).
\medskip

Having specified the ''true'' dynamics of $\phi$,
we ask now for its behaviour if the
observational resolution is such that $g_{00}^{\rm true}$
is replaced by its average (\ref{2.3}). Since we are interested
in the limit of large $\tau$, the damped oscillations of the
scale factors may be neglected. In Ref. \cite{FE2}, these
oscillations have been  studied in detail, and a
straightforward application of the insights gained there
reveals that $g^{-1/2}\partial_\tau g^{1/2}$ behaves
as $3/\tau + O(1/\tau^2)$ just as if one had used
$\hat{g}_{\rm eff}^{-1/2}\partial_\tau\hat{g}_{\rm eff}^{1/2}$
instead, where
$\hat{g}_{\rm eff}^{1/2} \equiv g_{\rm eff}^{1/2}
|g_{\rm eff}^{00}|^{1/2}$ (corresponding to the spatial part
of (\ref{2.4}). The only danger comes from $s$ at the
right hand side of (\ref{3.2}). Since the time scale
(\ref{2.2}) of the oscillations decreases to very small values,
the observed (coarse grained) average of $\phi$ will experience
some ''inertia''.
Neglecting the small-scale wiggles in $\phi$, one may just
replace $s$ by $<s>$ from (\ref{2.3}) in the version
(\ref{3.2}) of the field equation. However, this is
$-g_{00}^{\rm eff}$. Multiplying by $-g_{\rm eff}^{00}$, and
reshuffling terms, the resulting effective scalar field equation
is given by
\begin{equation}
- g_{\rm eff}^{-1/2} |g_{\rm eff}^{00}|^{1/2}
  \partial_\tau \,g_{\rm eff}^{1/2} |g_{\rm eff}^{00}|^{1/2}
  \partial_\tau \phi +
g_{\rm eff}^{-1/2}\partial_i \,g_{\rm eff}^{1/2}
g_{\rm eff}^{ij} \partial_j \phi
= V'(\phi).
\label{3.3}
\end{equation}
Thus, it becomes clear what has happened (to leading order in
$\tau^{-1}$): The original $g_{\rm true}^{00}$ has been replaced
by $g_{\rm eff}^{00}$, but such that part of it appears
already inside the $\tau$-derivative. In order to restore
a proper Laplacian (now with respect to the effective metric),
a correction term
$(\partial_\tau|g_{\rm eff}^{00}|^{1/2})|g_{\rm eff}^{00}|^{1/2}
\partial_\tau\phi \equiv (\Lambda/11)\tau\partial_\tau\phi$
(using $g_{\rm eff}^{00} = - \Lambda \tau^2/11$) is picked
up. Thus (\ref{3.3}) takes the form
\begin{equation}
\bigg(\Box_{\rm eff} + \frac{\Lambda}{11}\,\tau\partial_\tau\bigg)
\phi = V'(\phi).
\label{3.4}
\end{equation}
When expressed in terms of the effective cosmological proper time
$\eta$ as appearing in (\ref{2.6}), it may alternatively be
written as
\begin{equation}
\bigg(\Box_{\rm eff} + \sqrt{\frac{\Lambda}{11}}\,\partial_\eta\bigg)
\phi = V'(\phi).
\label{3.5}
\end{equation}
Hence, the effective scalar field equation contains -- in addition
to the effective Laplacian -- a first order time-derivative. Apart
from this, the overall structure is obtained from the ''true''
equation (\ref{3.1}) by simply replacing
$\Box_{\rm true} \rightarrow \Box_{\rm eff}$.
\medskip

The dynamical consequences of the additional term is estimated by
noting that the pure time-derivative contribution of
$\Box_{\rm eff}$ is
\begin{equation}
-\frac{\Lambda}{11}\bigg( \tau^2\partial_{\tau\tau} +
 4\tau\partial_\tau \bigg) \equiv
 - \bigg( \partial_{\eta\eta} + 3 \sqrt{\frac{\Lambda}{11}}\,
 \partial_\eta \bigg).
\label{3.6}
\end{equation}
The first derivatives act as damping terms (which prevent a
purely time-dependent $\phi\equiv\phi(\tau)$ from
rapidly ''rolling down'' the potential in most
inflationary models; see e.g. Ref. \cite{Brandenberger}). Comparison
with (\ref{3.4}) or (\ref{3.5}) shows that the correction terms
{\it increase} these damping effects, i.e. ensure that
$\phi$ varies a bit slower than it would without them.
The actual large-$\tau$ behaviour of $\phi$ is not so important
here, because one would anyway prefer to modify the model
such that inflation comes to end. However, what is really
interesting with (\ref{3.4}) or (\ref{3.5}) is its
structure. This will be considered in the next Section.

\section{The imprint of signature oscillations on matter}
\setcounter{equation}{0}

The effective scalar field equation contains the derivative
$\tau\partial_\tau$, hence a non-metric element. This is due
to the fact that the hypersurfaces $\tau = const$ -- although
just a coordinate to label space-time points from
an ''effective level'' geometrical point of view -- are
preferred: it is exactly at such hypersurfaces where
the ''microscopic'' signature changes occur. Equation
(\ref{3.4}) indicates how matter ''feels'' these
oscillations. Let us pose the question: What is the type of
non-metric structure one expects to play some role in
physical space-time? From the point of view we have adopted
(in order to obtain effective physics by some coarse graining
type of average), the only effective-level non-metric structures
of interest are the location of the hypersurfaces where
signature change occurs, and their ''density'' (which
is represented by $g_{00}^{\rm eff}$, i.e. by the quantity
$f$ defined in (\ref{2.7})). Hence, we expect $f$ to play
the role of a function on four- (or ten-) dimensional
space-time that is produced by the signature oscillations
and from which all their imprint on the effective physics
may be derived. This argument is not restricted to the very
simple FRW model we are considering. Hence, we expect
(in less symmetric situations) a scalar function
$f(x^\mu)$ to arise as a quantity encoding the location
of the signature change hypersurfaces ($f=const$) as well
as their density (presumably by its value). It is only
the equation (\ref{2.7}) that might be specific for our
model.
\medskip

Having identified $f$ as the ''trace'' of signature oscillations,
it is easy to recast the effective scalar field equation
(\ref{3.4}) into a generally covariant form. Using
$\partial_\tau f = -22/(\Lambda \tau^3)$ and
$\partial^\tau f \equiv g_{\rm eff}^{00}\partial_\tau f = 2/\tau$
(in general
$\partial^\mu f\equiv g_{\rm eff}^{\mu\nu}\partial_\nu f$),
one finds
\begin{equation}
\frac{\Lambda}{11}\, \tau\partial_\tau =
\frac{\partial^\mu f}{2 f} \,\partial_\mu.
\label{4.1}
\end{equation}
Hence, the final form of the scalar field equation is given by
\begin{equation}
\Big( \Box_{\rm eff} + \frac{1}{2} f^{-1}
          (\partial^\mu f) \partial_\mu \Big) \phi = V'(\phi).
\label{4.2}
\end{equation}
It states that $\phi$ {\it couples} in this particular
way to a background scalar field $f$. Since there is not much
chance for additional non-metric structure to arise,
more general types of matter fields are expected to pick up
additional couplings of this type as well (at least
to the lowest order in $\tau^{-1}$). Note that $f$ has
an ''absolute'' (coordinate invariant) meaning, and hence may be
called a scalar with respect to the effective geometry. In this
sense, the effective physics is perfectly covariant under
general coordinate transformations (as is (\ref{4.2})).
\medskip

In addition, $f$ carries dynamical information about the
''true'' background metric. This is just because it serves to
describe the hypersurfaces of signature change which
emerge from the (mixed-signature) Einstein field equations
in a dynamical way. However, it is not clear whether there is a
simple way to express this dynamical behaviour by means of
differential equations or a Lagrangian formulation
as a model for $(g_{\mu\nu}^{\rm eff},f)$ in less
symmetric situations. Let us just mention three observations
that might help exhibiting the nature of $f$.
\medskip

Firstly, as far as I can see, (\ref{4.2}) may not be obtained from
a local Lagrangian by variation of $\phi$, the reason being that
the operator $f^{-1}(\partial^\mu f)\partial_\mu$ is
formally anti-hermitean. This may be contrasted to
the case of a scalar in an external electromagnetic field
where one encounters $i A^\mu \partial_\mu$, the $i$
assuring formal hermiticity and thus allowing $A^\mu$ to be
coupled to a {\it complex} scalar.
\medskip

As a second observation we remark that in the purely
massive case ($V'(\phi)=m^2 \phi$) the f-terms in
the field equation (\ref{4.2}) may be absorbed to
lowest order into a rescaling of the field. Setting
\begin {equation}
\phi = - \,\frac{1}{4} f \psi,
\label{4.3}
\end{equation}
the field equation becomes
\begin{equation}
\Box_{\rm eff}\, \psi = \Big( m^2 + O(\tau^{-2}) \Big) \psi,
\label{4.4}
\end{equation}
without a $\partial_\mu \psi$ term. (This is somewhat analogous
to the above-mentioned electromagnetic case where the
identification $A_\mu \sim f^{-1} \partial_\mu f$ would
render $A_\mu$ gauge trivial. Then (\ref{4.3}) is the
analogue of a local U(1) gauge transformation). Hence, one
might try to consider rescaled quantities like $\psi$
(or even a conformally rotated metric) as the physical fields,
rather than $(\phi,g_{\mu\nu}^{\rm eff})$. To which extent
this is possible and reasonable deserves further study.
\medskip

The third observation concerns the status of $f$ ''by itself''
(i.e. without reference to $\phi$). A simple
calculation using (\ref{2.7}) and (\ref{3.6}) reveals that
\begin{equation}
\Box_{\rm eff} \,f = \frac{2 \Lambda}{11} f.
\label{4.5}
\end{equation}
Hence, at least in our simple model, $f$ satisfies the
equation of a massive scalar field with $m_f^2 = 2\Lambda/11$.
Due to the symmetries in the FRW-ansatz it is clear that only
the homogeneous mode is excited ($f \equiv f(\tau))$. It
would be interesting to know whether (\ref{4.5}) carries over
to less symmetric situations. (Recalling $f\equiv-g_{00}^{\rm eff}$,
(\ref{4.5}) looks like a coordinate condition emerging from
the proper-time nature of $\tau$ in the ''true'' geometry
and some dynamical input). However, in any case such a dynamical
equation would not apply without additional restrictions.
Even in the FRW model, the general purely time-dependent
solution to (\ref{4.5}) is $f(\tau) = c_1 \tau^{-1}+c_2 \tau^{-2}$,
whereas only its second part is realized.
\medskip

To summarize, we have provided some evidence that the scenario
of compactification by signature change leads to the appearance
of an effective, dynamical (''signaton'') scalar field $f$ that
couples to matter. In the FRW model we used for all explicit
computations, $f$ corresponds to the homogeneous mode of the
(standard) massive scalar field equation with
$m_f^2 = 2\Lambda/11$, although its dynamical nature in more
general cases is unclear. In the concluding Section we will
speculate on this last issue.

\section{Outlook}
\setcounter{equation}{0}

We have already mentioned that we expect the basic structure emerging
in our scenario to carry over to less symmetric situations. As in
all inflationary models, the FRW geometry may serve to approximate
physics locally. Globally, however, the geometry as well as the
topology may be completely different (see e.g. Ref. \cite{Linde}).
In this sense, it is conceivable to encounter a global
analogue $f(x^\mu)$ of the function $f(\tau)$ from (\ref{2.7}),
possibly subject to the massive scalar field equation (\ref{4.5})
everywhere or in some large portion of total space-time. Such
a ''signaton'' field would trace the global structure
of the hypersurfaces of signature change.
\medskip

It is thus of major importance for the scenario to find out
which dynamical laws the effective structures are
subject to, i.e. whether it is possible to formulate some effective
dynamics for the system $(g_{\mu\nu}^{\rm eff},f)$.
One step towards this direction would be to find out whether
$f$ is associated with some effective energy momentum tensor.
In the FRW model, the energy momentum tensor of the
four-dimensional part of the metric (\ref{2.4}),
is of the cosmological constant type
$T_{\mu\nu}^{\rm eff}= \Lambda_{\rm eff} g_{\mu\nu}^{\rm eff}$,
with $\Lambda_{\rm eff} = 3\Lambda/11$. (Here we have
neglected the curvature of the inflating ${\bf S}^3$,
which is compatible with all the large-$\tau$ limits
we have been considering in this paper).
Hence, any attempt to interpret this as (partially) arising
from $f$ would amount to set $T_{\mu\nu}(f) \sim g_{\mu\nu}^{\rm eff}$.
This is in turn not the standard scalar
field energy momentur tensor. One possibility to represent it
in terms of $f$ is an expression like
$T_{\mu\nu}(f) \sim -g_{\mu\nu}^{\rm eff}(\partial_\rho f)
(\partial^\rho f)/f^2$,
since
\begin{equation}
f^{-2} (\partial_\rho f) \partial^\rho f = -\, \frac{4 \Lambda}{11}
\label{4.6}
\end{equation}
is a constant.
Maybe in such an approach the field $f$ replaces the
cosmological constant completely at the effective level.
Another possibility would be to assign a potential $\sim \Lambda$
to $f$ and to note that the free part of the
standard scalar field energy momentum tensor when written in
terms of $f$
(namely $-(\partial_\mu f)(\partial_\nu f)+g_{\mu\nu}^{\rm eff}
(\partial_\rho f)(\partial^\rho f)/2$) is of higher
order $(\sim\tau^{-2})$ than the potential contribution
$(\sim \Lambda)$ and would not show up anyway in the
large-$\tau$ approximation. If this idea goes through, $f$
would be treated as a scalar field with constant potential.
However, as already stated in the preceeding Section,
$f$ will in any case be forced to obey additional restrictions,
and it may well turn out that it is in a rather strong way
more a ''function'' of $g_{\mu\nu}^{\rm eff}$ than a field
degree of freedom by its own.
In the FRW model, all these attempts coincide to leading
order in $\tau^{-1}$, and one would have to study more
general situations in order to distinguish between them.

\medskip
A further obstruction against $f$ behaving like an
ordinary scalar field is the fact that $f>0$. Furthermore,
it is dimensionless and can be made a
standard scalar only after a redefinition
$f_{\rm scalar} = \Lambda^{-1/4} f$.
However, there is some freedom in trying whether a function $F(f)$
fits better into a convenient dynamical scheme. This
may be seen by looking at the equations
\begin{equation}
\Box \ln f = \frac{6\Lambda}{11}, \qquad
\Box f^{3/2} = 0.
\label{4.7}
\end{equation}
Irrespective of these issues (that may contain aesthaetical
as well as physical aspects), one would like to know whether
$f$ or $F(f)$ develops some type of large-scale fluctuations or
even wave-like modes.
\medskip

As a last remark we add that it might be worth examining
whether the appearance of an effective ''signaton'' field
(giving rise to a preferred slicing!) has some
implications on the problem of time (and related issues)
when the ''classical signature change model'' is
quantized \cite{Martin}. Classical signature
change occurs when the potential appearing in the
Hamiltonian constraint (called $W$ in Ref. \cite{FE2})
is zero. Quantization of this model amounts to replace
$W \rightarrow |W|$ in the conventional Wheeler-DeWitt
equation \cite{FE1} (thus rendering the potential non-negative).
Within this framework, one could try to extract something
like $f$ on a full quantum or semi-classical level,
and then re-examine all the conceptual questions
encountered in quantum cosmology \cite{Isham}--\cite{AshtekarStachel}.


\begin{thebibliography}{00}

\bibitem{CremmerJulia}
{
 E. Cremmer and B. Julia,
 ''The SO(8) supergravity'',
 {\it Nucl. Phys. B} {\bf 159}, 141 (1979).
}

\bibitem{ChodosDetweiler}
{
 A. Chodos and S. Detweiler,
 ''Where has the fifth dimension gone?'',
{\it Phys. Rev. D} {\bf 21}, 2167 (1980).
}

\bibitem{FreundRubin}
{
 P. G. O. Freund and M. A. Rubin,
 ''Dynamics of dimensional reduction'',
 {\it Phys. Lett.} {\bf 97} B, 233 (1980).
}

\bibitem{Freund}
{
 P. G. O. Freund,
 ''Kaluza-Klein Cosmologies'',
 {\it Nucl. Phys. B} {\bf 209}, 146 (1982).
}

\bibitem{AppelquistChodos1}
{
 T. Appelquist and A. Chodos,
 ''Quantum Effects in Kaluza-Klein Theories'',
 {\it Phys. Rev. Lett.} {\bf 50}, 141 (1983).
}

\bibitem{AppelquistChodos2}
{
 T. Appelquist and A. Chodos,
 ''Quantum dynamics of Kaluza-Klein theories'',
 {\it Phys. Rev. D} {\bf 28}, 772 (1983).
}

\bibitem{AppelquistChodosMyers}
{
 T. Appelquist, A. Chodos and E. Myers,
 ''Quantum instability of dimensional reduction'',
 {\it Phys. Lett.} {\bf 127} B, 51 (1983).
}

\bibitem{RubinRoth}
{
 M. A. Rubin and B. D. Roth,
 ''Fermions and stability in five-dimensional Kaluza-Klein theory'',
 {\it Phys. Lett.} {\bf 127} B, 55 (1983).
}

\bibitem{ChodosMyers1}
{
 A. Chodos and E. Myers,
 ''Gravitational Contribution to the Casimir Energy in Kaluza-Klein
 Theories'',
 {\it Ann. Phys. (N.Y.)} {\bf 156}, 412 (1984).
}

\bibitem{ChodosMyers2}
{
 A. Chodos and E. Myers,
 ''Gravitational Casimir energy in non-Abelian Kaluza-Klein
 theories'',
 {\it Phys. Rev. D} {\bf 31}, 3064 (1985).
}

\bibitem{Sahdev1}
{
 D. Sahdev,
 ''Towards a realistic Kaluza-Klein cosmology'',
 {\it Phys. Lett.} {\bf 137} B, 155 (1984).
}

\bibitem{AbbottBarrEllis}
{
 R. B. Abbott, S. M. Barr and S. D. Ellis,
 ''Kaluza-Klein cosmologies and inflation'',
 {\it Phys. Rev. D} {\bf 30}, 720 (1984).
}

\bibitem{KolbLindleySeckel}
{
 E. W. Kolb, D. Lindley and D. Seckel,
 ''More dimensions -- Less entropy'',
 {\it Phys. Rev. D} {\bf 30}, 1205 (1984).
}

\bibitem{Sahdev2}
{
 D. Sahdev,
 ''Perfect fluid higher-dimensional cosmologies'',
 {\it Phys. Rev. D} {\bf 30}, 2495 (1984).
}

\bibitem{Yoshimura}
{
 M. Yoshimura,
 ''Effective action and cosmological evolution of scale factors
 in higher-dimensional curved space'',
 {\it Phys. Rev. D} {\bf 30}, 344 (1984).
}

\bibitem{MatznerMezzacappa}
{
 R. A. Matzner and A. Mezzacappa,
 ''Professor Wheeler and the Crack of Doom: Closed Cosmologies
 in the 5-d Kaluza-Klein Theory'',
 {\it Found. Phys.} {\bf 16}, 227 (1986).
}

\bibitem{Maeda}
{
 K. Maeda,
 ''Stability and attractor in a higher-dimensional cosmology:I,II'',
 {\it Class. Quantum Grav.} {\bf 3}, 233 (1986);
 {\it Class. Quantum Grav.} {\bf 3}, 651 (1986).
}

\bibitem{BailinLove}
{
 D. Bailin and A. Love,
 ''Kaluza-Klein theories'',
 {\it Rep. Prog. Phys.} {\bf 50}, 1087 (1987).
}

\bibitem{Wiltshire}
{
 D. L. Wiltshire,
 ''Global properties of Kaluza-Klein cosmologies'',
 {\it Phys. Rev. D} {\bf 36}, 1634 (1987).
}

\bibitem{SzydlowskiBiesiada}
{
 M. Szydlowski and M. Biesiada,
 ''Inflation as a dynamical effect of higher dimensions'',
 {\it Phys. Rev. D} {\bf 41}, 2487 (1990).
}

\bibitem{BertolamiKubyshinMourao}
{
 O. Bertolami, Yu. A. Kubyshin and J. M. Mour$\tilde{\rm a}$o,
 ''Stability of compactification in Einstein-Yang-Mills theories
 after inflation'',
 {\it Phys. Rev. D} {\bf 45}, 3405 (1992).
}

\bibitem{BertolamiMouraoKubyshin}
{
 O. Bertolami, J. M. Mour$\tilde{\rm a}$o and Yu. A. Kubyshin,
 ''On the stability of compactification after inflation'',
 {\it in}: H. Sato and T. Nakamura (eds.),
 {\it Proceedings of the 6th Marcel Grossmann
 Meeting 1991}, World Scientific (Singapore, 1992), p. 625.
}

\bibitem{ShafiWetterich1}
{
 Q. Shafi and C. Wetterich,
 ''Cosmology from higher-dimensional gravity''
 {\it Phys. Lett.} {\bf 129} B, 387 (1983).
}

\bibitem{ShafiWetterich2}
{
 Q. Shafi and C. Wetterich,
 ''Inflation with higher dimensional gravity'',
 {\it Phys. Lett.} {\bf 152}, 51 (1985).
}

\bibitem{ReuterWetterich}
{
 M.Reuter and C. Wetterich,
 ''Classical stability for spontaneous compactification
 in higher derivative gravity'',
 {\it Nucl. Phys. B} {\bf 289}, 757 (1987).
}

\bibitem{ShafiWetterich3}
{
 Q. Shafi and C. Wetterich,
 ''Inflation from higher dimensions'',
 {\it Nucl. Phys. B} {\bf 289}, 787 (1987).
}

\bibitem{Duff}
{
 M. J. Duff,
 ''Kaluza-Klein theory in perspective'',
 Talk delivered at the ''Oskar Klein Centenary Nobel Symposium'',
 Stockholm, September 1994, {\it preprint} hep-th/9410046.
}

\bibitem{FE2}
{
 F. Embacher,
 ''Signature change induces compactification'',
 {\it University Vienna preprint} UWThPh-1994-47
 (also available as gr-qc/9410012).
}

\bibitem{EllisSumeruketal}
{
 G. Ellis, A. Sumeruk, D. Coule and C. Hellaby,
 ''Change of signature in classical relativity'',
 {\it Class. Quantum Grav.} {\bf 9}, 1535 (1992).
}

\bibitem{Ellis}
{
 G. F. R. Ellis,
 ''Covariant Change of Signature in Classical Relativity'',
 {\it Gen. Relativ. Gravit.} {\bf 24}, 1047 (1992).
}

\bibitem{DereliTucker}
{
 T. Dereli and R. W. Tucker,
 ''Signature dynamics in general relativity'',
 {\it Class. Quantum Grav.} {\bf 10}, 365 (1993).
}

\bibitem{KernerMartin}
{
 R. Kerner and J. Martin,
 ''Change of signature and topology in a five-dimensional
 cosmological model'',
 {\it Class. Quantum Grav.} {\bf 10}, 2111 (1993).
}

\bibitem{DrayManogueTucker1}
{
 T. Dray, C. A. Manogue and R. W. Tucker,
 ''Particle Production from Signature Change'',
 {\it Gen. Relativ. Gravit.} {\bf 23}, 967 (1991).
}

\bibitem{DrayManogueTucker2}
{
 T. Dray, C. A. Manogue and R. W. Tucker,
 ''Scalar field equation in the presence of signature change'',
 {\it Phys. Rev. D} {\bf 48}, 2587 (1993).
}

\bibitem{HellabyDray}
{
 C. Hellaby and T. Dray,
 ''Failure of standard conservation laws at a classical change
 of signature'',
 {\it Phys. Rev. D} {\bf 49}, 5096 (1994).
}

\bibitem{DrayHellaby}
{
 T. Dray and C. Hellaby,
 ''The patchwork divergence theorem'',
 to appear in {\it J. Math. Phys}, {\it preprint} gr-qc/9404002.
}

\bibitem{GibbonsHartle}
{
 G. W. Gibbons and J. B. Hartle,
 ''Real tunneling geometries and the large-scale topology of
 the universe'',
 {\it Phys. Rev. D} {\bf 42}, 2458 (1990).
}

\bibitem{Hayward1}
{
 S. A. Hayward,
 ''Signature change in general relativity'',
 {\it Class. Quantum Grav.} {\bf 9}, 1851 (1992).
}

\bibitem{Hayward2}
{
 S. A. Hayward,
 ''On cosmological isotropy, quantum cosmology and the Weyl
 curvature hypothesis'',
 {\it Class. Quantum Grav.} {\bf 10}, L7 (1993).
}

\bibitem{FE1}
{
 F. Embacher,
 ''Comments on the multi-dimensional Wheeler-DeWitt equation'',
 Talk given at the International School-Seminar ''Multidimensional
 Gravity and Cosmology'', Yaroslavl, June 1994, to appear in
 the proceedings, {\it preprint} gr-qc/9409016.
}

\bibitem{Brandenberger}
{
 R. H. Brandenberger,
 ''Inflation and cosmic strings: Two mechanisms for producing
 structure in the universe'',
 {\it Int. J. Mod. Phys. A} {\bf 2}, 77 (1987).
}

\bibitem{Linde}
{
 A. Linde,
 ''Inflation and quantum cosmology'',
 {\it Physica Scripta T} {\bf 36}, 30 (1991).
}

\bibitem{Martin}
{
 J. Martin,
 ''Hamiltonian quantization of general relativity with the
 change of signature'',
 {\it Phys. Rev. D} {\bf 49}, 5086 (1994).
}

\bibitem{Isham}
{
 C. J. Isham,
 ''Canonical quantum gravity and the problem of time'',
 {\it in}: L. A. Ibort and M. A. Rodriguez (eds.),
 {\it Integrable systems, quantum groups, and quantum field theories},
 Kluwer Academic Publishers (London, 1993), 157 -- 287.
}

\bibitem{Kuchar}
{
 K. V. Kuchar,
 ''Time and interpretations of quantum gravity'',
 {\it in}: G. Kunstatter {\it et. al.} (eds.),
 {\it Proceedings of the 4th Canadian conference of general
 relativity and relativistic astrophysics},
 World Scientific (Singapore, 1992), 211 -- 314.
}

\bibitem{AshtekarStachel}
{
 A. Ashtekar and J. Stachel (eds.),
 {\it Conceptual Problems of Quantum Gravity},
 Birkh\"auser (Boston, 1991).
}

\end{thebibliography}
\end{document}